\documentclass[12pt]{iopart}
\usepackage{iopams} 
\usepackage{amstext,amsbsy,amsgen,amscd,amsopn,amsfonts,amssymb,amsbsy}
\usepackage{bm} 
\usepackage{graphicx} 
\usepackage{float} 
\usepackage{axodraw}
\usepackage{url}
\usepackage{hyperref}
\newcommand{\D}{{\rm d}}

\begin{document}
\paper{{\hfill\normalsize\tt DESY 06-046}}
\title[
Long-Lived Staus at Neutrino Telescopes]{
Long-Lived Staus at Neutrino Telescopes}
\author{Markus~Ahlers, J\"orn~Kersten, and Andreas~Ringwald}
\address{Deutsches Elektronen-Synchrotron DESY, Notkestra\ss e  85, 22607 Hamburg, Germany} 
\date{January 2006}

\begin{abstract}
We perform an exhaustive study of the role neutrino telescopes could play in the
discovery and exploration of supersymmetric extensions of the Standard Model
with a long-lived stau next-to-lightest superparticle. These staus are produced
in pairs by cosmic neutrino interactions in the Earth matter. We show that the
background of stau events to the standard muon signal is negligible and plays no
role in the determination of the cosmic neutrino flux. On the other hand, one
can expect up to 50 pair events per year in a cubic kilometer detector such as
IceCube, if the superpartner mass spectrum and the high-energy cosmic neutrino
flux are close to experimental bounds.
\end{abstract}
\ead{markus.ahlers@desy.de, joern.kersten@desy.de, andreas.ringwald@desy.de}

\maketitle

\section{Introduction}
Supersymmetry (SUSY) is currently the most popular extension of the Standard
Model (SM) and predicts a superpartner for each SM particle. In most models,
$R$-parity is conserved, so that the lightest superpartner (LSP) is stable. 
This makes it an excellent candidate for the dark matter.  Most studies assume
that this particle is a neutralino, which interacts weakly and may be observed 
therefore in direct dark matter searches. However, if SUSY is extended to
include gravity, there is an alternative LSP candidate, the gravitino.  As it is
the superpartner of the graviton, it takes part only in the gravitational
interaction.  Therefore, the decay of the next-to-lightest SUSY particle (NLSP)
is highly suppressed.\footnote{The same is true in theories with an axino LSP,
whose interactions are strongly suppressed by the large Peccei-Quinn scale, see
e.g.\ \cite{Brandenburg:2005he}. The NLSP can also be very long-lived if its mass
is very close to that of a neutralino LSP~\cite{Jittoh:2005pq}. 
We do not study these alternatives in
detail, but expect virtually the same results as in the case considered.} If it
is a charged particle such as the stau, $\tilde\tau_\text{R}$, the superpartner
of the right-handed tau, it can possibly be collected in collider experiments.  
Observation of the stau decays could then lead to an indirect discovery of the
gravitino.  This exciting possibility has attracted considerable interest
recently \cite{Buchmuller:2004rq,Hamaguchi:2004df,Feng:2004yi,Hamaguchi:2004ne,Brandenburg:2005he}. 

Within this scenario, we elaborate in this paper on an alternative experimental
approach in which the high energy necessary to produce SUSY particles is not
provided by man-made accelerators but by nature in the form of high energy
cosmic neutrinos.  If the stau NLSPs resulting from neutrino-nucleon
interactions inside the Earth are sufficiently long-lived, they can be detected
in large ice or water Cherenkov neutrino telescopes,  as pointed out in
\cite{Albuquerque:2003mi}. These observatories are designed to measure the flux
of high energy cosmic neutrinos via Cherenkov photons emitted by upward-going
muons and cascades, which are produced by weak interactions in the Earth. High
energy muons are visible if they originate up to a few tens of kilometers
outside the detector. This defines the effective detection volume, which is
limited by the energy loss of muons in matter. For a stau this effective
detection volume increases dramatically due to the much smaller energy loss in
matter~\cite{Reno:2005si}.  This might compensate the suppression of the
production cross section of SUSY particles compared to the one of the SM weak
interactions. Moreover, interactions of cosmic neutrinos with nucleons inside
the Earth will always produce pairs of staus, which appear as nearly parallel
muon-like tracks in the detector due to their large
boost-factor~\cite{Albuquerque:2003mi}. This is expected to be in contrast to
the SM processes, which lead to muon pairs only in rare cases.

The goal of this paper is to perform an exhaustive study of the role which
neutrino telescopes could play in the discovery and study of supersymmetric
extensions of the SM with a stau NLSP.  
The rate of stau pairs has already been estimated in Refs.~\cite{Albuquerque:2003mi} 
and~\cite{Bi:2004ys}. In this paper, we will focus on the limited detector response for 
stau events using the detailed calculation of the stau energy loss in Ref.~\cite{Reno:2005si}.
We will carefully calculate the NLSP
event rates for a cubic-kilometer neutrino observatory such as
IceCube~\cite{Ahrens:2002dv}, taking into account the dependence of the
detection efficiencies on stau energy and spatial resolution. 
For the
superpartner mass spectrum, we will use, for illustration, the benchmark point
corresponding to SPS~7 from \cite{Allanach:2002nj} and a toy model where all
superpartner masses are just above the experimental limits. Similarly, for the
yet unknown high energy cosmic neutrino flux we will adopt as a benchmark the
Waxman-Bahcall flux~\cite{Waxman:1998yy}, which assumes a simultaneous
production of cosmic neutrinos, protons, and gamma rays in astrophysical
accelerator sources of the observed cosmic rays with comparable luminosities.

The outline of the paper is as follows. In section~\ref{NLSPPROD}, we calculate
the differential production cross section of SUSY particles. We discuss their
lifetime and energy loss in the Earth in section~\ref{NLSPPROG}. Finally, we
compute the resulting flux of staus and the event rates at neutrino observatories in
section~\ref{DET}. 

\section{NLSP Production}\label{NLSPPROD}

Inelastic scattering on nucleons  is the dominant interaction process of the
high energy cosmic neutrinos in the atmosphere or the Earth.  In this section,
we calculate the differential cross sections for neutrino-quark scattering into
sleptons and squarks. They determine the energy spectrum of the stau NLSPs. 
The resulting SUSY contribution to the neutrino-nucleon cross section is
compared with the corresponding SM contribution.

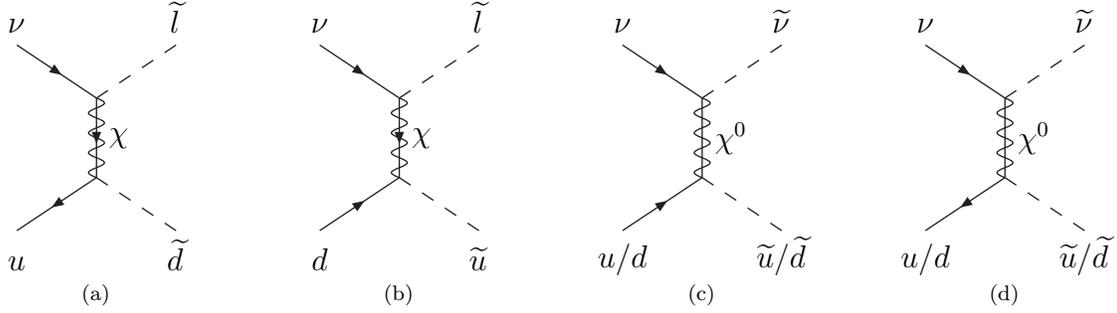
\begin{figure}[t]
\begin{minipage}{0.23\linewidth}\centering
\begin{picture}(80,105)(10,3)
\ArrowLine(50,40)(20,20)
\ArrowLine(20,90)(50,70)
\ArrowLine(50,70)(50,40)
\Photon(50,40)(50,70){3}{4}
\DashLine(50,40)(80,20){5}
\DashLine(50,70)(80,90){5}
\Text(20,5)[b]{$u$}
\Text(20,95)[b]{$\nu$}
\Text(80,5)[b]{${\widetilde d}$}
\Text(80,95)[b]{${\widetilde l}$}
\Text(55,55)[l]{$\chi$}
\end{picture}\\ \scriptsize(a)
\end{minipage}
\hfill
\begin{minipage}{0.23\linewidth}\centering
\begin{picture}(80,105)(10,3)
\ArrowLine(20,20)(50,40)
\ArrowLine(20,90)(50,70)
\ArrowLine(50,70)(50,40)
\Photon(50,40)(50,70){3}{4}
\DashLine(50,40)(80,20){5}
\DashLine(50,70)(80,90){5}
\Text(20,5)[b]{$d$}
\Text(20,95)[b]{$\nu$}
\Text(80,5)[b]{${\widetilde u}$}
\Text(80,95)[b]{${\widetilde l}$}
\Text(55,55)[l]{$\chi$}
\end{picture}\\ \scriptsize(b)
\end{minipage}
\hfill
\begin{minipage}{0.23\linewidth}\centering
\begin{picture}(80,105)(10,3)
\ArrowLine(20,20)(50,40)
\ArrowLine(20,90)(50,70)
\Line(50,70)(50,40)
\Photon(50,40)(50,70){3}{4}
\DashLine(50,40)(80,20){5}
\DashLine(50,70)(80,90){5}
\Text(20,3)[b]{$u/d$}
\Text(20,95)[b]{$\nu$}
\Text(80,3)[b]{$\widetilde u/\widetilde d$}
\Text(80,95)[b]{$\widetilde \nu$}
\Text(55,55)[l]{$\chi^0$}
\end{picture}\\ \scriptsize(c)
\end{minipage}
\hfill
\begin{minipage}[c]{0.23\linewidth}\centering
\begin{picture}(80,105)(10,3)
\ArrowLine(50,40)(20,20)
\ArrowLine(20,90)(50,70)
\Line(50,70)(50,40)
\Photon(50,40)(50,70){3}{4}
\DashLine(50,40)(80,20){5}
\DashLine(50,70)(80,90){5}\Text(20,3)[b]{$u/d$}
\Text(20,95)[b]{$\nu$}
\Text(80,3)[b]{${\widetilde u}/{\widetilde d}$}
\Text(80,95)[b]{${\widetilde \nu}$}
\Text(55,55)[l]{$\chi^0$}
\end{picture}\\ \scriptsize(d)
\end{minipage}
\caption[]{Chargino and neutralino exchange diagrams}\label{diagrams}
\end{figure}

The parton-level SUSY contributions to charged and neutral current interactions between neutrinos
and nucleons are the chargino and neutralino exchange diagrams shown in
Fig.~\ref{diagrams},
as well as analogous interactions with the heavier quarks.
The reactions produce sleptons and squarks, which promptly decay into the
lighter stau, usually composed predominantly of the superpartner of the
right-handed tau. The parton-level cross sections for these
diagrams\footnote{The cross sections for the analogous processes with
anti-neutrinos are the same.} are given by
\numparts\begin{eqnarray}
\fl\qquad\frac{\D\sigma^{(a)}_{\nu\bar u}}{\D t} &=& 
\frac{\pi\alpha^2}{2s_W^4}\frac{1}{s^2}(tu-m_{\tilde l_L}^2m_{\tilde q}^2)
\left[\frac{Z_-^{1i}Z_-^{1i}}{(t-m_{\chi_i}^2)}\right]^2,
\label{inta}\\
\fl\qquad\frac{\D\sigma^{(b)}_{\nu d}}{\D t} &=& 
\frac{\pi\alpha^2}{2s_W^4}\frac{1}{s}
\left[\frac{m_{\chi_i}Z_+^{1i}Z_-^{1i}}{(t-m_{\chi_i}^2)}\right]^2,
\label{intb}\\
\fl\qquad\frac{\D\sigma^{(c)}_{\nu(u/d)}}{\D t} &=&
\frac{\pi\alpha^2}{2s_W^4}\frac{4}{c_W^4}\frac{1}{s^2}\left\lbrace
(tu-m_{\tilde l_L}^2m_{\tilde q}^2) 
\left[\frac{{{\cal Y}^i}_{\!\!\!u_R/d_R}{{\cal Y}^i}_{\!\!\!\nu}}{(t-m_{\chi_i^0}^2)}\right]^2+s
\left[\frac{m_{\chi_i^0}{{\cal Y}^i}_{\!\!\!u_L/d_L} {{\cal Y}^i}_{\!\!\!\nu}}{(t-m_{\chi_i^0}^2)}
\right]^2\right\rbrace,\label{intc}\\ 
\fl\qquad\frac{\D\sigma^{(d)}_{\nu(\bar u/\bar d)}}{\D t} &=&
\frac{\pi\alpha^2}{2s_W^4}\frac{4}{c_W^4}\frac{1}{s^2}\left\lbrace
(tu-m_{\tilde l_L}^2m_{\tilde q}^2) 
\left[\frac{{{\cal Y}^i}_{\!\!\!u_L/d_L}{{\cal Y}^i}_{\!\!\!\nu}}{(t-m_{\chi_i^0}^2)}\right]^2+s
\left[\frac{m_{\chi_i^0}{{\cal Y}^i}_{\!\!\!u_R/d_R} {{\cal Y}^i}_{\!\!\!\nu}}{(t-m_{\chi_i^0}^2)}
\right]^2\right\rbrace.\label{intd}
\end{eqnarray}\endnumparts
Summation over repeated indices is implied. We use the conventions and notations
of \cite{Rosiek:1989rs,Denner:1992vz}. The masses of the four neutralinos and
two charginos are denoted by $m_{\chi_i^0}$ and $m_{\chi_i}$, respectively.  The
mixing matrices are $Z_N$ and $Z_\pm$. The neutralino couplings ${\cal
Y}^i = YZ_N^{1j}s_W+T^3Z_N^{2j}c_W$ depend on the hypercharge $Y=Q-T^3$ and the
weak isospin $T^3$:
\begin{equation*}
\renewcommand{\arraystretch}{1.5}
\begin{array}{llll}
{\cal Y}_{u_L}^i &=  \frac{1}{6}Z_N^{1i}s_W+\frac{1}{2}Z_N^{2i}c_W,
\qquad&{\cal Y}_{d_L}^i &= \frac{1}{6}Z_N^{1i}s_W-\frac{1}{2}Z_N^{2i}c_W,\\
{\cal Y}_{u_R}^i &= \frac{2}{3}Z_N^{1i}s_W,
\qquad&{\cal Y}_{d_R}^i &= -\frac{1}{3}Z_N^{1i}s_W,\\
{\cal Y}_{\nu}^i &= -\frac{1}{2}Z_N^{1i}s_W+\frac{1}{2}Z_N^{2i}c_W.&&
\end{array}
\end{equation*}
In the following we will focus on two SUSY mass spectra. One is given by the
benchmark point corresponding to SPS~7~\cite{Allanach:2002nj} for a
gauge-mediated SUSY breaking (GMSB) scheme with a messenger mass of 80~TeV. 
\begin{figure}[t]
\begin{minipage}{\linewidth}\centering
\includegraphics[width=0.60\linewidth]{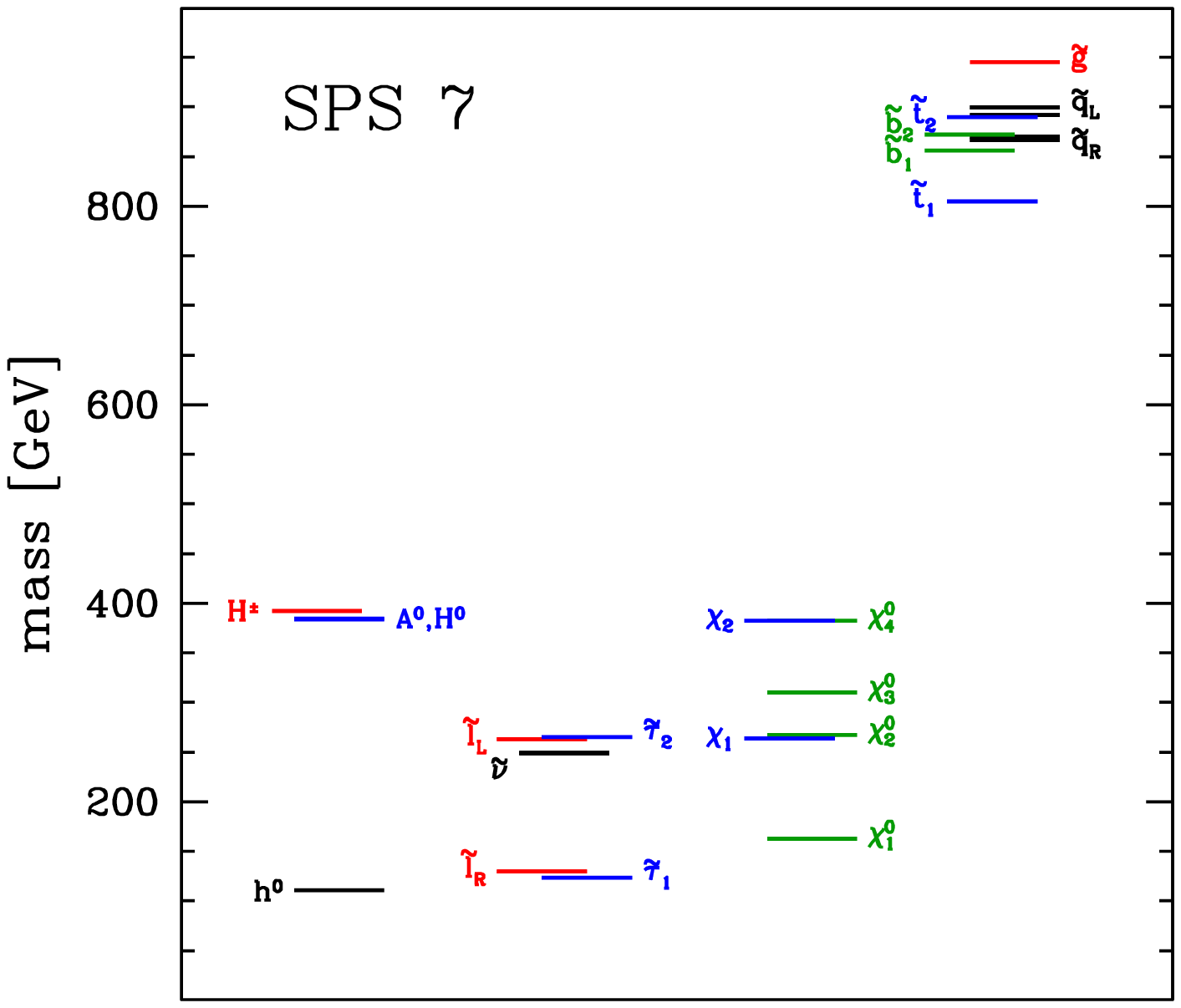} 
\end{minipage}
\caption[]{The SUSY mass spectrum of the benchmark point corresponding to 
SPS~7~\cite{Allanach:2002nj}.}\label{SPS7}
\end{figure}
\begin{figure}[t]
\centering
  \includegraphics[width=0.70\linewidth]{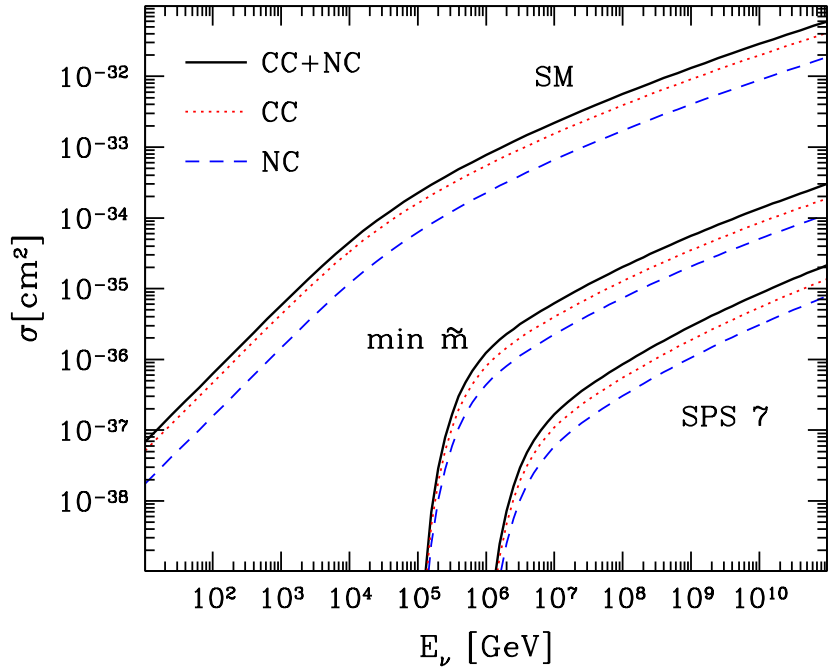}
\caption{The charged (red dotted) and neutral (blue dashed) current cross section for 
neutrino-nucleon scattering compared to the chargino and neutralino exchange for the SPS~7 
benchmark point and a scenario with light squark masses of 300 GeV.}
\label{sigma}
\end{figure}
The corresponding mass spectrum calculated by {\sc SoftSusy
2.0.4}~\cite{Allanach:2001kg} is shown in Fig.~\ref{SPS7}. The other spectrum
(denoted by ``min $\widetilde m$'' in the following) consists of light
charginos, neutralinos and sleptons at 100~GeV and squarks at 300~GeV. It is not
motivated by any particular SUSY breaking scenario, but oriented at current
experimental limits, in order to give an impression of what can be obtained in a
very optimistic scenario. 

The resulting neutrino-nucleon cross sections are shown in Fig.~\ref{sigma},
compared to the SM contribution from $W$ and $Z$ exchange.
Here and in the following computations we have used the {\sc CTEQ6D} parton
distribution functions~\cite{Pumplin:2002vw}. The contribution to the cross sections from
charged currents is about twice as large as the one from neutral currents.

In Eqs.~(\ref{inta})--(\ref{intd}), we have neglected family mixings and
contributions proportional to Yukawa couplings.  We have taken into account the
exchange of the heavier neutralinos and charginos, since generically their
contribution to the cross sections is large and can even dominate.  At the SPS~7
benchmark point we consider, assuming equal neutralino masses turns out to be a
rather good approximation that helps simplify analytical calculations. 
Nevertheless, we use the exact expressions in the numerical calculations
discussed in the following.

\section{NLSP Propagation}\label{NLSPPROG}

The squarks and sleptons produced in the interactions discussed in the previous
section promptly decay into stau NLSPs. Now, we discuss the survival probability
of the latter after propagation towards the Cherenkov light detector. The
corresponding effective detection volume is limited by energy loss and decay.  

\subsection{Energy Loss}

In the case of the muon, the mean energy loss per column depth $z$, measured in
$\text{g}\,\text{cm}^{-2}$, is given as
\begin{equation}
-\left\langle\frac{\D E_\mu}{\D z}\right\rangle \approx \alpha_\mu + \beta_\mu E_\mu.\label{EL}
\end{equation}
Here, $\alpha_\mu$ is determined by ionization effects and $\beta_\mu$ accounts
for bremsstrahlung, pair-production, and photohadronic processes. In general,
these coefficients are weakly energy dependent. For our purpose, it is
sufficient to approximate their values by constants, 
$\alpha_\mu\approx2\cdot10^{-3}\ \text{GeV}\,\text{cm}^2\,\text{g}^{-1}$ and
$\beta_\mu\approx4\cdot10^{-6}\ \text{cm}^2\,\text{g}^{-1}$. For muons above a
critical energy $E_\mu^\text{cr} = \alpha_\mu/\beta_\mu \approx 500$~GeV,
radiative energy losses dominate, which are proportional to the energy itself.
Hence, Eq.~(\ref{EL}) can be used to determine the muon energy if $E_\mu\gg
E^\text{cr}_\mu$. We will take this critical energy as an effective cut-off in
the following since the detection efficiency becomes very small
below~\cite{Ahrens:2003ix}. The range of a muon is then given as
\begin{equation}
R(E_\mu,E_\mu^\text{cr}) \approx 
\frac{1}{\beta_\mu}\ln\left(\frac{\alpha_\mu+\beta_\mu E_\mu}{\alpha_\mu+\beta_\mu E_\mu^\text{cr}}\right).
\end{equation}
In the case of a stau, the radiative term in Eq.~(\ref{EL}) is suppressed, since
$\beta_{\widetilde\tau}$ satisfies $\beta_{\widetilde\tau}m_{\widetilde\tau}
\approx \beta_\mu m_\mu$. At large energies this gives an increase of the stau
range compared to that of the muon by the mass ratio $m_{\widetilde\tau}/m_\mu$.
In the following we will use the results of~\cite{Reno:2005si} for the range of
the stau.\footnote{If the stau NLSP has a large admixture from $\widetilde\tau_L$, 
the superpartner of the left-handed tau, charged and neutral current effects, which have 
not been considered in~\cite{Reno:2005si}, might be important~\cite{BEACOM}. 
Assuming a small admixture of $\lesssim10$\%, as it is the case at the SPS~7 benchmark 
point, we expect that these effects can safely be neglected.
}
 
\subsection{Decay}

The lifetime of the stau NLSP can be very long, since its decay into the
gravitino LSP can proceed only gravitationally. The corresponding decay length
$L$ for relativistic staus, in units of the Earth's diameter $2 R_\oplus$, 
is (e.g.~\cite{Giudice:1998bp}) 
\begin{equation}\label{LofNLSP}
\bigg(\frac{L}{2 R_\oplus}\bigg) 
\approx \bigg(\frac{m_{\tilde \tau}}{100\ \text{GeV}}\bigg)^{-6}
\bigg(\frac{m_{3/2}}{400\ \text{keV}}\bigg)^2\bigg(\frac{E_{\tilde \tau}}{500\ \text{GeV}}\bigg).
\end{equation}
The minimal stau energy we consider is about 500~GeV, corresponding to the
critical energy of the muons.\footnote{Note, however, that the critical energy
of the stau is given as $E^\text{cr}_{\widetilde\tau} \approx
(m_{\widetilde\tau}/m_\mu) E^\text{cr}_\mu$, according to
$\beta_{\widetilde\tau} \approx (m_\mu/m_{\widetilde\tau})\beta_\mu$.} Hence,
staus with a mass not much larger than 100~GeV will always reach the detector,
if the gravitino is heavier than 400~keV.\footnote{Actually, we expect that
our results remain unchanged even if the gravitino is lighter by an order of
magnitude, since the dominant contribution to the event rate is due to staus
with energies considerably larger than 500~GeV.} In this case, the gravitino is a viable
candidate for the dark matter \cite{Pagels:1981ke}.  Constraints from big bang
nucleosynthesis and the cosmic microwave background yield an upper limit on the
gravitino mass between 10 and 100~GeV \cite{Cerdeno:2005eu}.  Masses in the
lower part of the allowed region are typical for models with gauge-mediated SUSY
breaking, while masses of some tens of GeV can occur in gravity and gaugino
mediation (for a review, cf.\ e.g.\ \cite{Chung:2003fi}).

\section{Detection Rate at Neutrino Telescopes}\label{DET}

There are several Cherenkov light based neutrino telescopes currently operating
or under construction in the Antarctic
(AMANDA~\cite{Andres:1999hm,Ahrens:2003pv} and IceCube~\cite{Ahrens:2002dv}), in Lake Baikal~\cite{Belolaptikov:1997ry,Balkanov:1999rf}, and in the
Mediterranean (ANTARES~\cite{Aslanides:1999vq}, NEMO~\cite{Piattelli:2005hz},
and NESTOR~\cite{Grieder:2001ha}). The total flux of staus through a detector is
proportional to the initial flux of high energy cosmic neutrinos $F$, attenuated
inside the Earth according to the total inelastic neutrino-nucleon cross section
$\sigma_{\nu N}$. A slepton 
(squark) produced with an energy $E_{\widetilde l}$
($E_{\widetilde q}$) and a cross section $\sigma^{\rm SUSY}_{\nu N}$ will
promptly decay into a stau which looses energy by radiative processes according
to Eq.~(\ref{EL}) and reaches the detector with energy $E_{\widetilde\tau}$. 
The flux of staus
per area $A$, time $t$, steradian $\Omega$, and energy $E_{\widetilde\tau}$ is then given as
\begin{equation}
\fl\qquad\frac{\D^4 N}{\D t\, \D A\, \D \Omega\, \D E_{\widetilde\tau}} \approx 
\int\limits_{0}^{z_\text{tot}}\frac{\D z}{m_\text{p}}
\int\limits_{E_\text{min}}^{E_\text{max}}\!\! \D E_{\nu}\sum\limits_{i = \widetilde l,\widetilde q}\, 
\left|\frac{\D E_i}{\D E_{\widetilde\tau}}\right|
\frac{\D \sigma^{\rm\scriptscriptstyle SUSY}_{\nu N}}{\D E_i} \exp\left(-\frac{z}{m_\text{p}}\sigma_{\nu N}
\right)F(E_\nu) .\label{fluxofstaus}
\end{equation}
Here, $m_\text{p}$ is the proton mass, and the column depth $z$ of the Earth is
related to $l$, the distance to the detector in the line of sight, as $-\D
z=\rho(l,\theta)\D l$. For the calculation of the total column depth of the
Earth we use the density profile given in \cite{Gandhi:1995tf} and a detector center at 1.9 km depth, as it is
the case for IceCube.  
The measure $|\D
E_i/\D E_{\widetilde\tau}|$ takes into account the energy loss of a stau as well
as the mean energy fraction it carries in the decay of the initial slepton or
squark. For the SPS~7 benchmark point we have estimated $\langle
E_{\widetilde\tau}\rangle \approx 0.3 E_{\widetilde q}$ and $\langle
E_{\widetilde\tau}\rangle \approx 0.5 E_{\widetilde l}$, respectively. For the
optimistic scenario with light sparticles this changes to $\langle
E_{\widetilde\tau}\rangle \approx 0.5 E_{\widetilde q}$ and $\langle
E_{\widetilde\tau}\rangle \approx E_{\widetilde l}$, respectively.

For the flux of high energy cosmic neutrinos, we adopt the Waxman-Bahcall (WB)
flux, $E_\nu^2F(E_\nu) \approx 2\cdot 10^{-8}\ \text{cm}^{-2}\,\text{s}^{-1}\,\text{sr}^{-1}\,\text{GeV}$ per
flavor~\cite{Waxman:1998yy}. This flux is based on the assumption that the
observed ultra high energy cosmic rays are protons from extragalactic
astrophysical acceleration sites. In these violent environments protons,
neutrinos, and gamma rays are produced with comparable luminosities.     As a
maximal neutrino energy we take $10^{10}$~GeV. We have checked that an increase
of this cut-off does not affect our results.

\begin{figure}[t]
\begin{minipage}[t]{\linewidth}
\begin{center}
\includegraphics[width=0.48\linewidth]{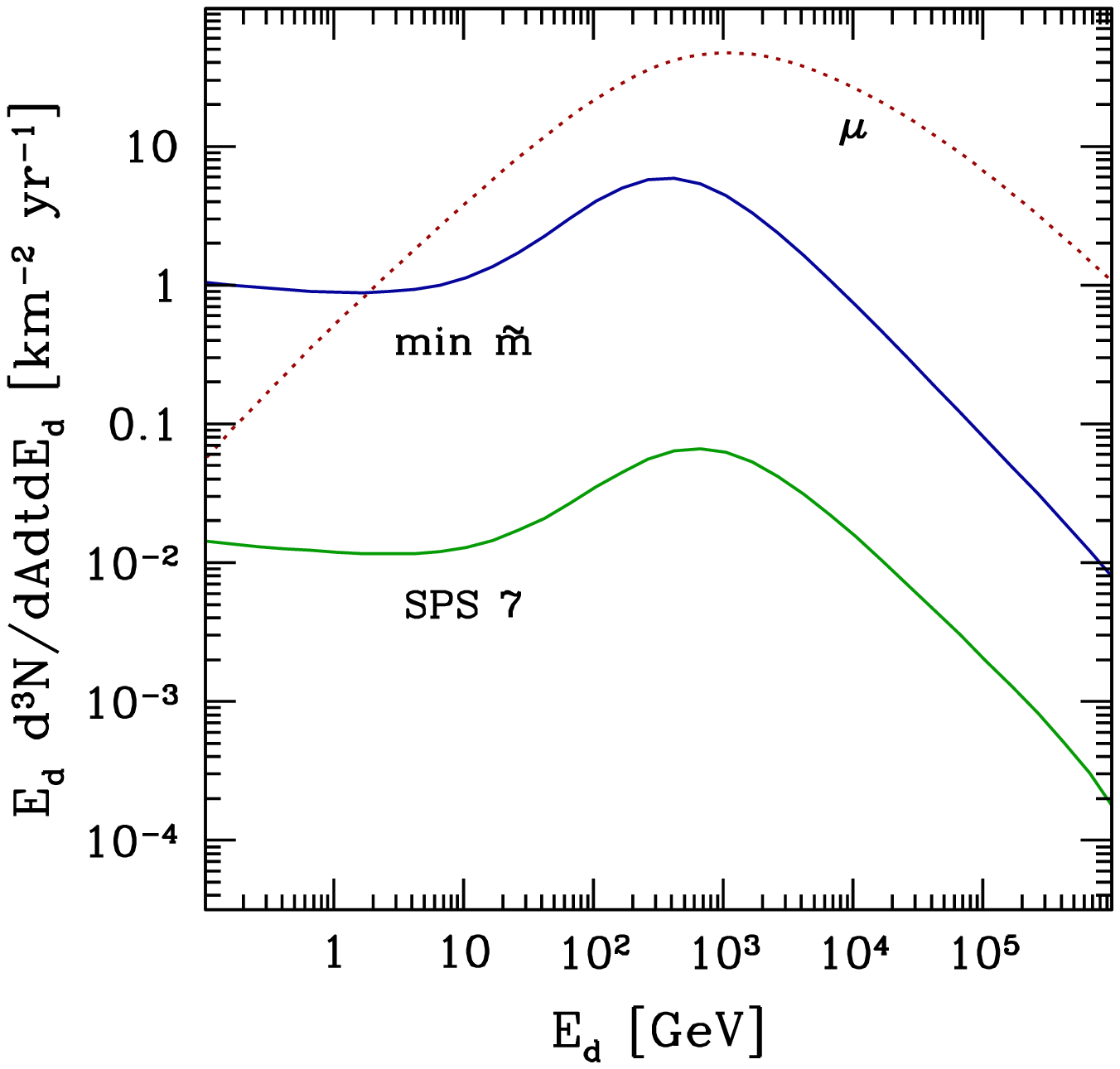}
\hfill
 \includegraphics[width=0.48\linewidth]{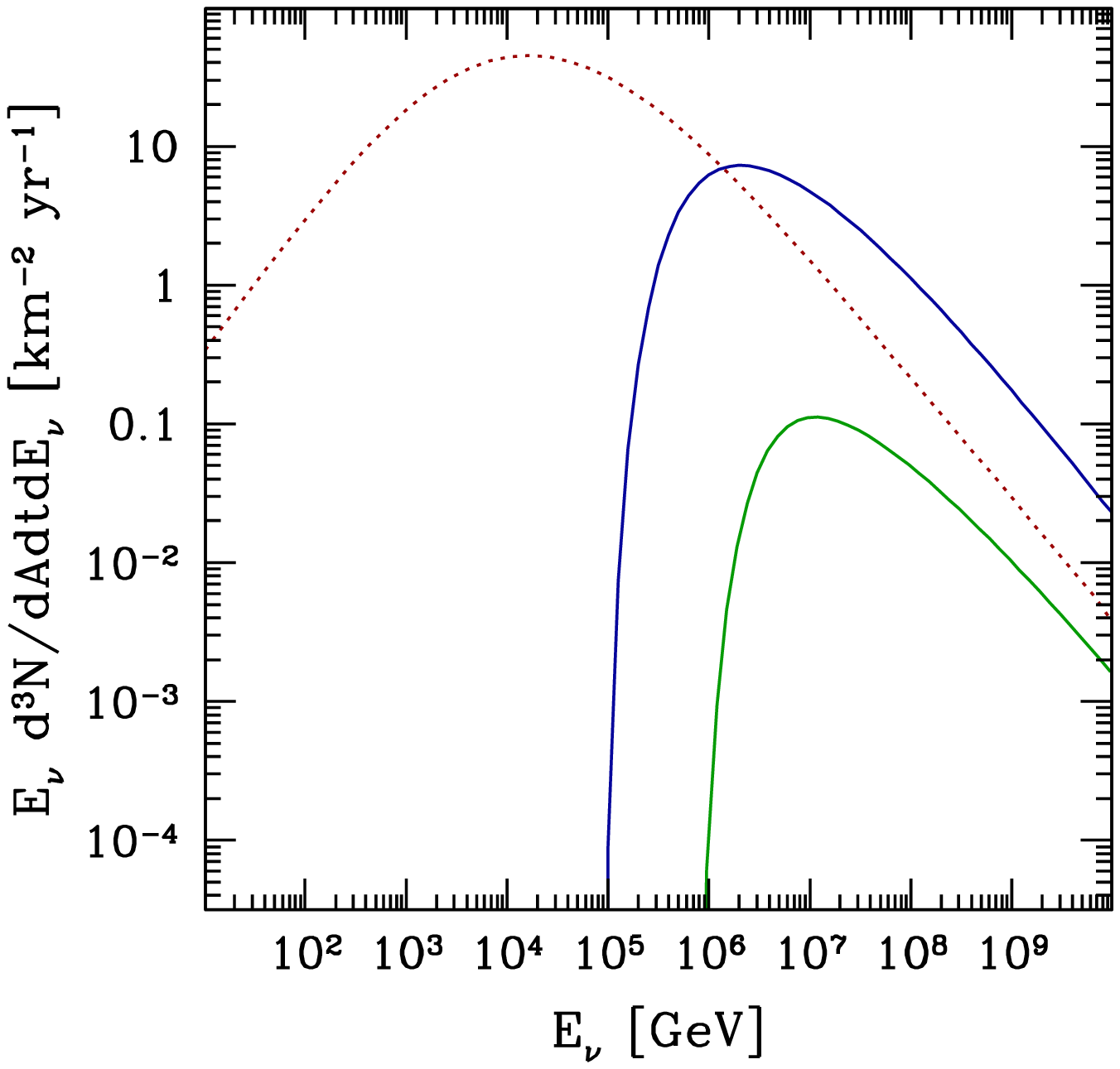}
\end{center}
\end{minipage}
\\[0.5cm]
\begin{minipage}[t]{\linewidth}
\begin{center}
\includegraphics[width=0.48\linewidth]{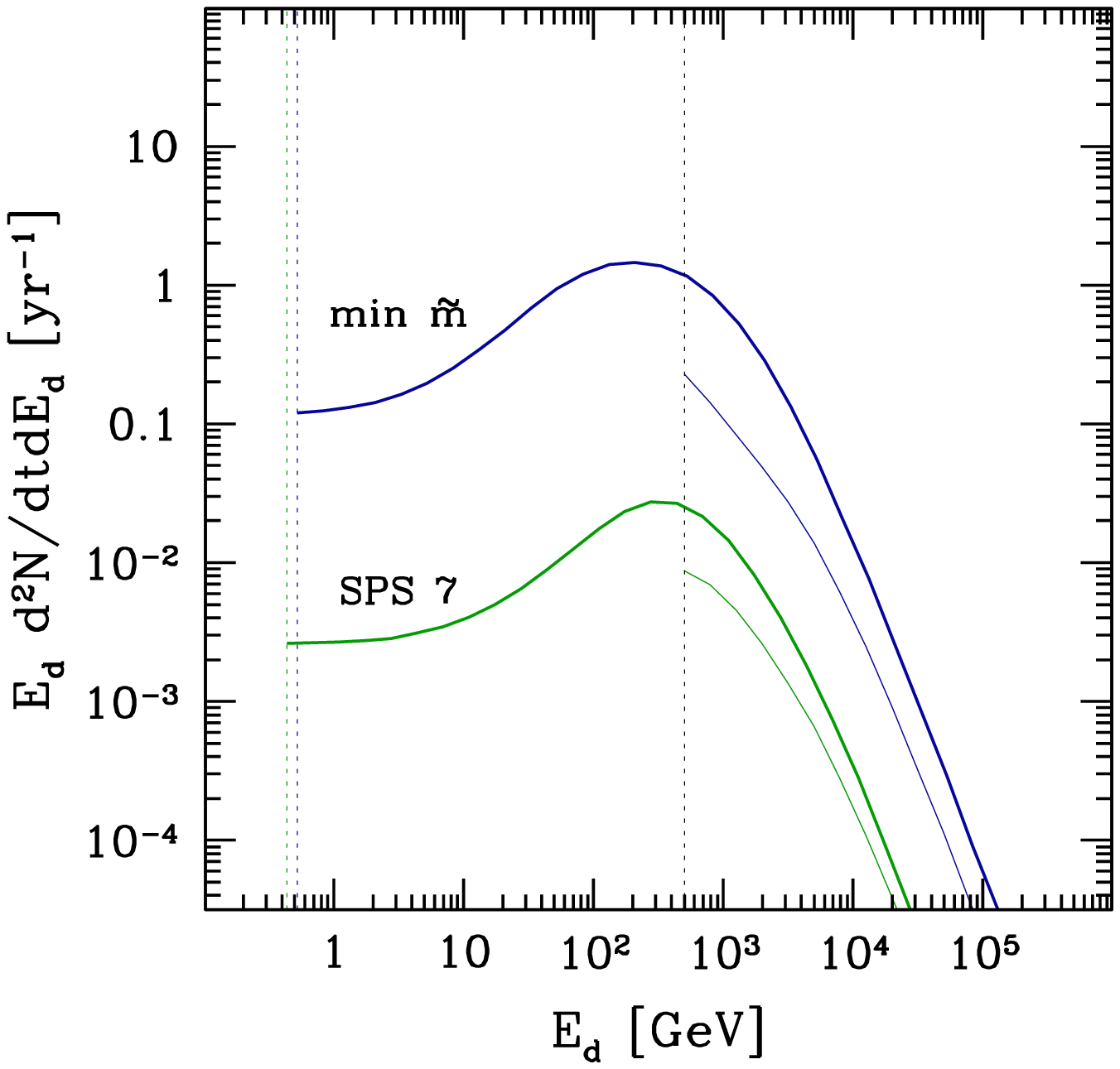}
\hfill
\includegraphics[width=0.48\linewidth]{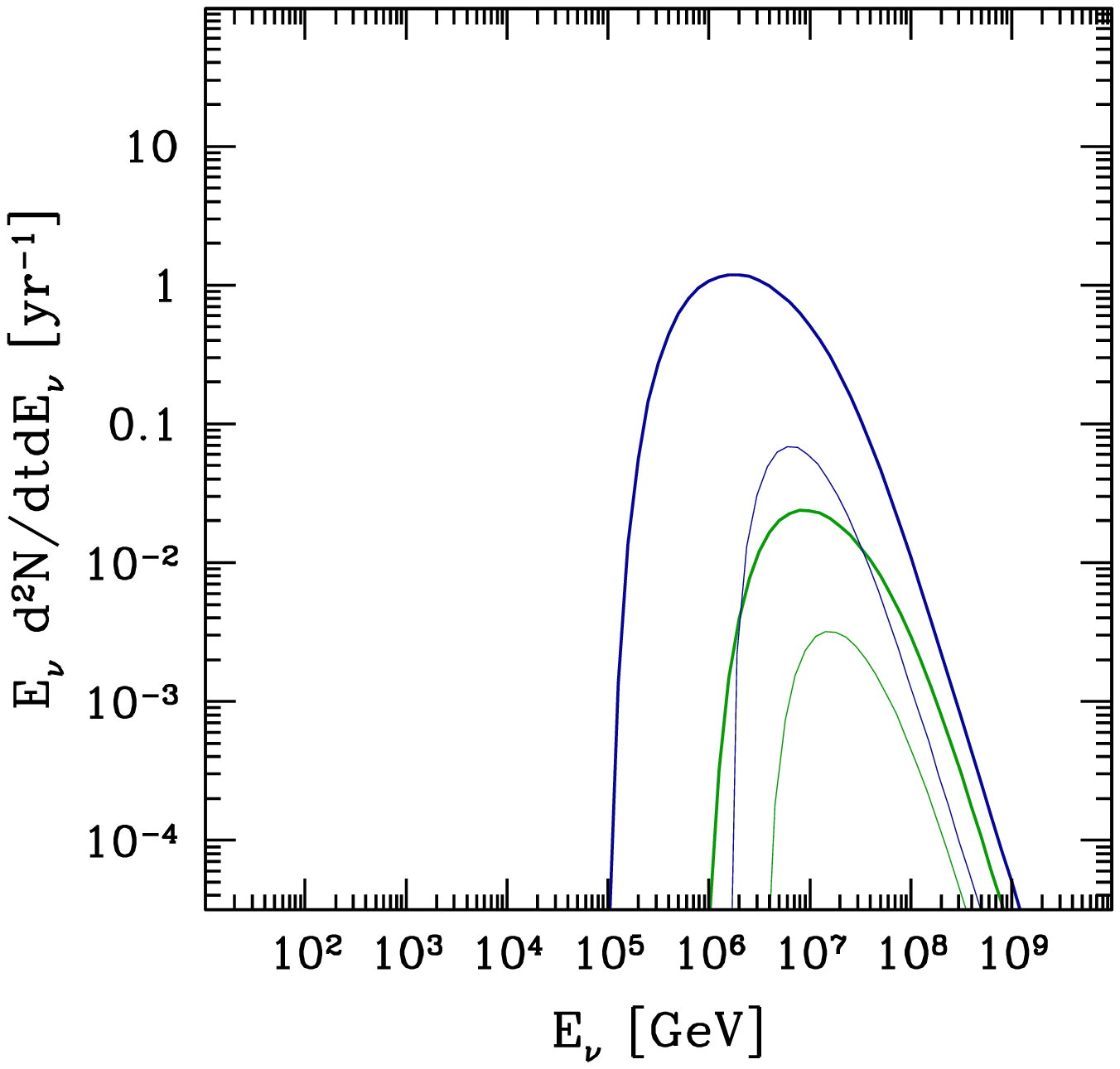}
\end{center}
\end{minipage}
\caption[]{Expected fluxes and event rates for the example of IceCube and the Waxman-Bahcall 
neutrino flux \cite{Waxman:1998yy}. The results are shown in terms of the detected energy 
$E_\text{d}$ (left panels) and the neutrino energy $E_\nu$ (right panels).
We have used a SUSY mass spectrum  
with light squarks of 300 GeV (blue upper solid) as well as the one from the benchmark point corresponding 
to SPS~7 
(green lower solid).
\\ 
{\bf Upper panels}: Fluxes of upward-going muons (red dotted)  
and staus (solid). 
For optimistic SUSY scenarios the spectrum of staus may dominate over the flux of muons 
at large neutrino energies (right panel). However, the detected spectrum of staus will be subdominant 
(left panel) due to the detector calibration explained in the text.    
\\ 
{\bf Lower panels}: Rates of parallel-stau events for an energy cut-off at 
$E^\text{cr}_{\mu}=\alpha_{\mu}/\beta_{\mu}$ (thick solid) and at 
$E^\text{cr}_{\widetilde\tau}=\alpha_{\widetilde\tau}/\beta_{\widetilde\tau} = 
(m_{\widetilde\tau}/m_{\mu})E^\text{cr}_{\mu}$ (thin solid), respectively.}\label{rates}
\end{figure}

The analysis of particle tracks observed in neutrino telescopes is calibrated
for muons. Their energy $E_\mu$ can be reconstructed by the energy loss $\Delta
E$ per column depth $\Delta z$ and Eq.~(\ref{EL}). This is appropriate for muons
with energies above the critical energy
$E^\text{cr}_\mu=\alpha_\mu/\beta_\mu\approx 500\ \text{GeV}$.
A high-energy stau with energy $E_{\widetilde\tau}$ will deposit a much smaller
energy fraction in the detector compared to a muon with the same energy, due to
the reduced radiative energy loss. This has two effects. Firstly, the critical
energy of a stau $E_{\widetilde\tau}^\text{cr} =
\alpha_{\widetilde\tau}/\beta_{\widetilde\tau}\approx(m_{\widetilde\tau}
/m_\mu)E^\text{cr}_{\mu}$ is much higher compared to the muons. Secondly, due to
the calibration of the detector, high-energy staus will be detected as muons with
reduced energy $E_\text{d} \approx
(m_\mu/m_{\widetilde\tau})E_{\widetilde\tau}$. Note that this correctly
identifies the boost factor of the particle $\gamma =
E_{\widetilde\tau}/m_{\widetilde\tau} = E_\text{d}/m_\mu$. Also the Cherenkov
angle is consistently identified by $\cos\theta_C \propto
(1-1/\gamma^{2})^{-1/2}$. For that reason, we will present our results
in Fig.~\ref{rates} in terms of the detected
energy $E_\text{d}$. Of course, the total number of events does not change by this recalibration. 
Note that the critical energy for the stau
has the same value for muons and staus in terms of the detected energy.  
For completeness, we show our results also in terms of the neutrino
energy.

\subsection{One-Muon vs.\ One-Stau Events}

The upper panels of Fig.~\ref{rates} show the differential fluxes of muons and staus through the IceCube 
detector. 
The calculation of the total rate of events requires the knowledge of the detection efficiency of
the corresponding particle depending on its energy and direction. This is
usually combined with the cross sectional area of the detector as an effective
area $A_\text{eff}$. The IceCube Collaboration has determined the effective area
$A^\mu_\text{eff}(\cos\theta,E_\mu)$ for upward-going muons by Monte-Carlo simulations in
\cite{Ahrens:2003ix}. The angle-averaged values are shown in Tab.~\ref{absrates}
together with the expected rates of muon events  
for the WB flux. The expected rates are in good 
agreement with those quoted in \cite{Ahrens:2003ix}, if one takes into account the different 
normalizations of the initial neutrino fluxes. 

Despite the fact that the staus are produced in pairs, they may also produce one-stau events, 
for example if one particle of the pair misses the detector. These are indistinguishable from one-muon
events, at least for high energies.  
A very conservative upper bound on the rate of one-stau events is given by the total rate of staus
computed from Eq.~(\ref{fluxofstaus}) and a detection efficiency 
$A^{\widetilde\tau}_\text{eff}(E_{\widetilde\tau}) \approx A^\mu_\text{eff}(E_\text{d})$. 
This upper bound is also shown in Tab.~\ref{absrates} and 
Fig.~\ref{rates} for the two different SUSY mass spectra.  
It shows that in the fiducial energy region one-stau events are subdominant compared to the flux of
muons, even for the most optimistic case of very light squarks with masses
around 300~GeV. This seems to be generic, independent of the initial high-energy
cosmic neutrino flux, as we have checked. Hence, the total rate of one-particle events
can be used for reconstructing the neutrino flux without taking into account the
contributions from NLSPs. This decouples the analysis of the extra-galactic neutrino flux from
the search for parallel-stau events, to which we turn now.

\begin{table}[t]
\begin{indented}
\item[]
\renewcommand{\arraystretch}{1.2}
\vspace{0.5cm}
\begin{tabular*}{\linewidth}{@{}l@{\extracolsep\fill}ccccc@{}}
\hline
$E_\text{d}$ [GeV]&$(\,10\, ,\,10^2\,)$&$(\,10^2\, ,\,10^3\,)$&$(\,10^3\, ,\,10^4\,)$&$(\,10^4\, ,\,10^5\,)$
&$(\,10^5\, ,10^6\,)$\\
$A^\mu_{\rm eff}$ [${\rm km}^2$]&$0.1$&$0.6$&$0.8$&$1.0$&$1.2$\\
$N_{\mu}$&3&60&100&80&20\\
\hline
$N_{\widetilde\tau}$ (minimal $\widetilde m$)&$<0.9$&$<10$&$<6$&$<0.9$&$<0.1$\\
$N_{\widetilde\tau}$ (SPS 7)&$<0.009$&$<0.1$&$<0.1$&$<0.02$&$<0.003$\\
\hline
\end{tabular*}
\end{indented}
\caption[]{The expected number of muon and stau events for a one year observation at IceCube, assuming the 
Waxman-Bahcall neutrino flux per flavor,  
$E_\nu^2F(E_\nu) \approx 2\cdot 10^{-8}\ \text{cm}^{-2}\,\text{s}^{-1}\,\text{sr}^{-1}\,\text{GeV}$ 
\cite{Waxman:1998yy}. The effective areas of IceCube have been calculated by Monte-Carlo simulations and 
are given in \cite{Ahrens:2003ix}.}\label{absrates}
\end{table}

\subsection{Rate of Parallel-Stau Events}

The detection efficiency of stau pairs, i.e.\ coincident parallel tracks, 
depends on the energies and directions of the staus, as in the case of
single events, and also on their separation. As a detailed Monte-Carlo simulation
is beyond the scope of this study, we will use a simple approximation for the resulting 
angle-averaged effective area in the following. Firstly, we introduce a low energy cut-off for the detectable 
stau energy. As examples, we use the critical energies of the muon and of the stau. Secondly, we
assume that the observatory can identify an event of coincident parallel
stau tracks, if they are separated by more than 50~m~\cite{SPIERING} and less than 1~km. Here, we
approximate the opening angle by the angle between sleptons and squarks in the
laboratory frame. If the pair event survives these cuts, we assume an effective area of 
$1\ \text{km}^2$ for its detection. 

The lower panels of Fig.~\ref{rates} 
show the expected rates of these pair events for the two SUSY scenarios. Since the staus
of a pair have different energies, 
we show the average of the corresponding spectra. 
Note that a higher energy threshold in the detector does also affect the range of the second stau. 
This has an effect on the pair rates even if one stau is above the threshold as can be seen in the plots. 
The main reason for the  suppression at large energies $E_\text{d}$ is the requirement that 
the opening angle of the staus be compatible with the restriction on the separation.
On the other hand, the distance of the interaction point from the detector has
to be smaller than the corresponding range of the second stau. This mainly
causes the reduction at small energies $E_\text{d}$. For the Waxman-Bahcall
neutrino flux and one year of observation at IceCube, we predict about 0.09 pair
events for the benchmark point corresponding to SPS~7 and  about 5 events for
our toy model with light sparticle masses. This calculation assumes an effective
low-energy cut-off at $E^\text{cr}_\mu$. 
We also show the effect of increasing the cut-off to 
$E^\text{cr}_{\widetilde\tau}= (m_{\widetilde\tau}/m_\mu)E^\text{cr}_\mu$. This
might be necessary due to the reduced radiative energy loss of the stau. Our
expectations are then reduced to 0.009 and 0.2 events, respectively.
Currently, the IceCube Collaboration quotes an upper limit at $90\%$ confidence level of
$E_\nu^2F(E_\nu) \lesssim 3\cdot 10^{-7}\
\text{cm}^{-2}\,\text{s}^{-1}\,\text{sr}^{-1}\,\text{GeV}$ per flavor for the
diffuse flux of cosmic neutrinos~\cite{Ackermann:2005aa}. This is approximately
one order of magnitude above the Waxman-Bahcall flux taken for our calculation.
If the cosmic neutrino flux does not lie much below this limit, 
the predicted event rates will increase by a
factor 10, giving up to 50 pair events for the optimistic SUSY mass
spectrum. 

The expected background of parallel muon pair events from random coincidences is bounded from above by the 
number of muons arriving within a (generous) time-window of $1\ \mu\text{s}$. Even this is several orders of 
magnitude below the stau pair event rate,
\begin{equation}
\displaystyle\frac{N_{\mu+\mu}(E)}{N
_\mu(E)} \lesssim \frac{1}{2}\text{min}\Bigg(1\,;{1\ \mu\text{s}} \cdot 
\frac{\D N_\mu}{\D t}\Bigg)\sim\mathcal{O}(10^{-12}).
\end{equation}
However, double muon events from processes like $\nu_\mu+ N
\rightarrow\mu^- +\pi^+ +X\rightarrow\mu^- +\mu^++\nu_{{\mu}}+X$ are expected to be more
likely. We expect that these contributions are still subdominant compared to stau pairs due to 
meson-nucleon interactions in the Earth. This should be studied in more detail in the future.

\section{Summary and Discussion}\label{SD}

High energy cosmic neutrinos collide with matter in the Earth at center of mass
energies beyond the capability of any earth-bound experiment. The attempt to
measure the cosmic ray fluxes in various observatories is therefore tightly
connected to extrapolations of the SM interactions of these particles to very
high energies. Besides, one can look for deviations from the SM predictions for
the strength of the interactions or even for new particles.

In this paper, we have examined the question whether neutrino telescopes could
play a significant role in the discovery and study of supersymmetric extensions
of the SM with a gravitino LSP and a long-lived stau NLSP. In these models,
neutrino-nucleon interactions can produce pairs of staus which show up as
parallel tracks in the detector~\cite{Albuquerque:2003mi}.

We have argued that events with a single stau in the detector are virtually
indistinguishable from muon events at high energies.  However, the reduced
energy loss of staus in matter, which increases the effective volume, reduces
also the detection efficiency in the telescope.  As a result, single stau events
play only a subdominant role compared to muon events.  Consequently, the
reconstruction of the initial high-energy neutrino flux from the total rate of
muon-like events can proceed assuming SM interactions alone. 

If the superpartner masses are close to their experimental limits and if the
cosmic neutrino flux is also close to its current experimental bound, we expect
up to 50 pair events per year in a cubic kilometer detector such as IceCube,
with negligible background.  Less favorable event rates are obtained for a less
optimistic cut-off in the stau energy or for SUSY mass spectra of commonly used
scenarios for SUSY breaking, mainly because the squarks are significantly
heavier.

If the gravitino mass is larger than 400~keV, as we have assumed, the stau decay
length is larger than the Earth's diameter.  Thus, the spectrum is independent
of the gravitino mass.  For lighter gravitinos, the rate of stau pairs will drop
starting at the low-energy end of the spectrum.  Therefore, it might be possible
to obtain information about the gravitino mass, if the spectrum is measured very
accurately and if the superpartner masses are known.  However, with the small
number of pair events we find, this appears challenging. 

\section*{Acknowledgements}

We would like to thank John Beacom, Daniel Garcia Figueroa, Steen Hannestad, Koichi
Hamaguchi, Rolf Nahnhauer, J\"urgen Reuter, Tania Robens, Christian Spiering,
Frank Steffen, and Peter Zerwas for discussions. MA would like to thank the
members of the IceCube Collaboration at Zeuthen for their hospitality.
This work has been supported by the ``Impuls- und Vernetzungsfonds'' of the
Helmholtz Association, contract number VH-NG-006.


\section*{References}
\frenchspacing
\bibliography{refs}
\bibliographystyle{h-physrev3}
\end{document}